\documentclass[12pt]{article}

   \usepackage{graphicx}
   \usepackage{amsfonts}
   \usepackage{amssymb}
   \usepackage{amsthm}
   \usepackage{amsmath}

   \usepackage[active]{srcltx}

   \usepackage{here}
   \usepackage{cite}
   \usepackage{color}
   \usepackage{soul}

   \usepackage[affil-it]{authblk}

   \input paperdef
   \graphicspath{{figs/}}

    \newcommand{\Abc}{The}
    \newcommand{\abc}{the}
    \newcommand{\xyz}{and}

\begin{document}

\thispagestyle{empty}

\title{
\vspace*{-2cm}
The LHC Higgs Boson Discovery: Updated implications for Finite
Unified Theories and the SUSY breaking scale}
\date{}
\author{ \hspace*{7mm} S. Heinemeyer$^{1,2,3}$\thanks{email: Sven.Heinemeyer@cern.ch} , M. Mondrag\'on$^4$\thanks{email: myriam@fisica.unam.mx} ,
G. Patellis$^5$\thanks{email: patellis@central.ntua.gr} ,
N. Tracas$^5$\thanks{email: ntrac@central.ntua.gr}$\,$  \hspace*{19mm} and G. Zoupanos$^{5,6}$\thanks{email: George.Zoupanos@cern.ch}\\
{\small
$^1$Instituto de F\'{\i}sica Te\'{o}rica, Universidad Aut\'{o}noma de Madrid Cantoblanco, 28049 Madrid, Spain\\
$^2$Campus of International Excellence UAM+CSIC, 
Cantoblanco, 28049, Madrid, Spain \\
$^3$Instituto de F\'{\i}sica de Cantabria (CSIC-UC),  E-39005 Santander, Spain \\
$^4$Instituto de F\'{\i}sica,  Universidad Nacional
Aut\'onoma de M\'exico,  A.P. 20-364, M\'exico 01000\\ %
%$^4$ Theoretical Physics Department, CERN, Geneva, Switzerland\\%
$^5$ Physics Department,   Nat. Technical University, 157 80 Zografou, Athens, Greece\\
$^6$ Max-Planck Institut f\"ur Physik, F\"ohringer Ring 6, D-80805
  M\"unchen, Germany
}
}

\maketitle

\vspace{-1.0cm}

\abstract{Finite Unified Theories (FUTs) are $N=1$ supersymmetric Grand Unified
Theories which can be made finite to all orders in perturbation theory,
based on the principle of reduction of couplings. The latter consists in
searching for renormalization group invariant relations among parameters
of a renormalizable theory holding to all orders in perturbation theory.
FUTs have proven very successful so far. In particular, they predicted the
top quark mass one and half years before its experimental discovery, while
around five years before the Higgs boson discovery a particular FUT
was predicting the light Higgs boson in the mass range $\sim 121-126 ~{\rm GeV}$,
in striking agreement with the discovery at LHC.
Here we review the basic properties of the supersymmetric theories and in
particular finite theories resulting from the application of the method
of reduction of couplings in their dimensionless and dimensionful sectors.
Then we analyse the phenomenologically favoured FUT, based on
SU(5). This particular FUT leads to a finiteness constrained version of
the MSSM, which naturally predicts a relatively heavy spectrum with
coloured supersymmetric particles above 2.7 TeV, consistent with the
non-observation of those particles at the LHC. The electroweak
supersymmetric spectrum starts below 1 TeV and large parts of the allowed
spectrum of the lighter might be accessible at CLIC. The FCC-hh will be
able to fully test the predicted spectrum.}

\vspace{0.2cm}

\noindent
{\tt 
IFT-UAM/CSIC-18-013
}

%The method of reduction of couplings is applied to a Finite Unified Theory.
%We search for renormalization group invariant relations among couplings of a renormalizable theory which
%holds to all orders in perturbation theory. The method leads to relations, at the unification
%scale, between gauge and Yukawa couplings (in the dimensionless sectors of the theory) and relations %among
%the couplings of the trilinear terms and the Yukawa couplings and similarly among the scalar masses and %the
%gaugino mass satisfying moreover a sum rule (in the soft breaking sector).

\vfill
\newpage

\section{Introduction}
In 2012 the discovery of a new particle at the LHC was
announced~\cite{Aad:2012tfa,Chatrchyan:2012ufa}. Within theoretical and
experimental uncertainties the new particle is compatible with predictions
for the Higgs boson of the Standard Model
(SM)~\cite{ATLAS:2013mma,Chatrchyan:2013lba}, constituting a milestone 
in the quest for understanding the physics of electroweak symmetry breaking
(EWSB). However, taking the experimental results and the respective 
uncertainties into account, also many models beyond the SM can accomodate the
data. Furthermore, the hierarchy problem, the neutrino masses, the Dark
Matter, the over twenty free parameters of the model, just to name some
questions, ask for a more fundamental theory to answer some, if not all, of
those.  

Therefore, one of the main aims of this fundamental theory is to relate these free parameters, or rephrasing it,
to achieve a {\it reduction of these parameters} in favour of a smaller number (or ideally only one).
This reduction is usually based in the introduction of a larger symmetry, rendering the theory more predictive.
Very good examples are the Grand Unified Theories (GUTs) \cite{Pati:1973rp,Georgi:1974sy,Georgi:1974yf,Fritzsch:1974nn,Carlson:1975gu} and their supersymmetric extensions \cite{Dimopoulos:1981zb,Sakai:1981gr}.
The case of minimal $SU(5)$ is one example, where the number of couplings is reduced to one due to the
corresponding unification. Data from LEP \cite{Amaldi:1991cn} suggested that a
$N=1$ global supersymmetry (SUSY)
 \cite{Dimopoulos:1981zb,Sakai:1981gr} is required in order for the prediction to be viable.
Relations among the Yukawa couplings are also suggested in GUTs. For example, the $SU(5)$
predicts the ratio of the tau to the bottom mass $M_{\tau}/M_b$ \cite{Buras:1977yy} in the SM.
GUTs introduce, however, new complications such as the different ways of breaking this larger symmetry
as well as new degrees of freedom.

A way to relate the Yukawa and the gauge sector, in other words achieving Gauge - Yukawa Unification (GYU)
\cite{Kubo:1995cg,Kubo:1997fi,Kobayashi:1999pn} seems to be a natural extension of the GUTs.
The possibility that $N=2$ SUSY \cite{Fayet:1978ig} plays such a role is
highly limited due to the prediction of light mirror fermions. Other
phenomenological drawbacks appear in composite models and superstring 
theories.

A complementary approach is to search for all-loop Renormalization Group Invariant (RGI) relations
\cite{Zimmermann:1984sx,Oehme:1984yy} which hold below the Planck scale and are preserved down to the unification scale
\cite{Kapetanakis:1992vx,Mondragon:1993tw,Kubo:1994bj,Kubo:1994xa,Kubo:1995hm,Kubo:1995cg,Kubo:1996js,Kubo:1997fi,Kobayashi:1999pn}.
With this approach  Gauge - Yukawa unification (GYU) is possible. 
A remarkable point is that, assuming finiteness at one-loop in $N = 1$ gauge theories, RGI relations that guarantee finiteness to all orders in perturbation theory can be found \cite{Lucchesi:1987he,Piguet:1986td,Lucchesi:1996ir}.

The above approach seems to need SUSY as an essential ingredient.
However the breaking of SUSY has to be understood too, since it provides the SM with several predictions for its free parameters.
Actually, the RGI relation searches have been extended
to the soft SUSY breaking (SSB) sector \cite{Kubo:1996js,Jack:1995gm,Zimmermann:2001pq,Mondragon:2013aea} relating parameters of mass dimension one and two. This is indeed possible to be done on the RGI surface which is defined by the solution of the reduction equations.

Applying the reduction of couplings method to $N=1$ SUSY theories has led to
very interesting phenomenological developments. Previously, an appealing
``universal'' set of soft scalar masses was assumed in the SSB sector of SUSY
theories, given that, apart from economy and simplicity, (1) they are part of
the constraints that preserve finiteness up to two loops
\cite{Jones:1984cu,Jack:1994kd}, (2) they are RGI up to two loops in more
general SUSY gauge theories, subject to the condition known as $P=1/3Q$
\cite{Jack:1995gm} and (3) they appear in the attractive dilaton dominated
SUSY breaking superstring scenarios
\cite{Ibanez:1992hc,Kaplunovsky:1993rd,Brignole:1993dj}. However, further
studies exhibited problems all due to the restrictive nature of the
``universality'' assumption for the scalar masses. For instance, (a) in Finite
Unified Theories (FUTs) the universality predicts that the lightest SUSY
particle is a charged particle, namely the superpartner of the $\tau$ lepton
$\tilde{\tau}$, (b) the MSSM with universal soft scalar masses is inconsistent
with the attractive radiative electroweak symmetry breaking and, worst of all,
(c) the universal soft scalar masses lead to charge and/or colour breaking
minima deeper than the standard vacuum \cite{Casas:1996wj}. Therefore, there
have been attempts to relax this constraint without loosing its attractive
features. First, an interesting observation was made that in $N=1$ GYU
theories there exists a RGI sum rule for the soft scalar masses at lower
orders; at one loop for the non-finite case \cite{Kawamura:1997cw} and at two
loops for the finite case \cite{Kobayashi:1997qx}. The sum rule manages to
overcome the above unpleasant phenomenological consequences. Moreover, it was
proven \cite{Kobayashi:1998jq} that the sum rule for the soft scalar masses is
RGI to all orders for both the general and the finite case. Finally, the exact
$\beta$-function for the soft scalar masses in the
Novikov-Shifman-Vainstein-Zakharov (NSVZ) scheme
\cite{Novikov:1983ee,Novikov:1985rd,Shifman:1996iy} for the softly broken SUSY
QCD has been obtained \cite{Kobayashi:1998jq}. The use of RGI both in the
dimensionful and dimensionless sector, together with the above mentioned sum
rule, allows for the construction of realistic and predictive $N=1$ all-loop
finite $SU(5)$ SUSY GUTs, also with interesting predictions, as was shown
in \citeres{Kapetanakis:1992vx} and  \cite{Kubo:1995cg,Kubo:1994xa,Kubo:1995zg,Heinemeyer:2007tz,Mondragon:2013aea,Heinemeyer:2012yj,Heinemeyer:2013nza,Heinemeyer:2014mya}.

This paper is organised as follows. In Section~2 we review
the theoretical basis of the method of the reduction of
couplings, which is extended in the subsection~2.1 to the dimensionful
parameters. Section~3 is devoted to Finiteness in the dimensionless sector
of a SUSY theory in some detail. In Section~4 we discuss
the implications of the method of reduction of couplings in the
SUSY breaking sector of an $N=1$ SUSY theory including
the finite case. Then in Section~5 we review the best Finite Unified
Model selected previously on the basis of agreement with the known
experimental data at the time~\cite{Heinemeyer:2007tz}.
The current set-up of experimental constraints and predictions is briefly
reviewed in Section~6 and applied to our best Finite Unified Model in
Section~7, including in particular the latest improvements in the prediction
of the light Higgs boson mass (as implemented in {\tt FeynHiggs}). 
Our conclusions can be found in Section~8.

%%%%%%%%%%%%%%%%%%%%%%%%%%%%%%%%%%%%%%%%%%%%%%%%%%%%%%%%%%%%%%%%%%%%%%%%%%%%%%%
%%%%%%%%%%%%%%%%%%%%%%%%%%%%%%%%%%%%%%%%%%%%%%%%%%%%%%%%%%%%%%%%%%%%%%%%%%%%%%%

\section{Theoretical basis}
In this section we outline \abc\ idea of reduction of couplings.
Any RGI relation among couplings
(which does not depend on \abc\ renormalization
scale $\mu$ explicitly) can be expressed,
in \abc\ implicit form $\Phi (g_1,\cdots,g_A) ~=~\mbox{const.}$,
which
has to satisfy \abc\ partial differential equation (PDE)
\beq
\mu\,\frac{d \Phi}{d \mu} = {\vec \nabla}\cdot {\vec \beta} ~=~
\sum_{a=1}^{A}
\,\beta_{a}\,\frac{\partial \Phi}{\partial g_{a}}~=~0~,
\eeq
where $\beta_a$ is \abc\ $\beta$-function of $g_a$.
This PDE is equivalent
to a set of ordinary differential equations,
the so-called reduction equations (REs) \cite{Zimmermann:1984sx,Oehme:1984yy,Oehme:1985jy},
\beq
\beta_{g} \,\frac{d g_{a}}{d g} =\beta_{a}~,~a=1,\cdots,A~,
\label{redeq}
\eeq
where $g$ \xyz\ $\beta_{g}$ are \abc\ primary
coupling \xyz\ its $\beta$-function,
and \abc\ counting on $a$ does not include $g$.
Since maximally ($A-1$) independent
RGI ``constraints''
in \abc\ $A$-dimensional space of couplings
can be imposed by \abc\ $\Phi_a$'s, one could in principle
express all \abc\ couplings in terms of
a single coupling $g$.
However, a closer look to \abc\ set of \refeqs{redeq} reveals that their
general solutions contain as many integration constants as \abc\ number of
equations themselves. Thus, using such integration constants we have just
traded an integration constant for each ordinary renormalized coupling,
and consequently, these general solutions cannot be considered as
reduced ones. \Abc\ crucial requirement in \abc\ search for RGE relations is
to demand power series solutions to \abc\ REs,
\beq
g_{a} = \sum_{n}\rho_{a}^{(n)}\,g^{2n+1}~,
\label{powerser}
\eeq
which preserve perturbative renormalizability.
Such an ansatz fixes \abc\ corresponding integration constant in each of
the REs \xyz\ picks up a special solution out of \abc\ general one.
Remarkably, \abc\ uniqueness of such power series solutions can be
decided already at \abc\ one-loop level
\cite{Zimmermann:1984sx,Oehme:1984yy,Oehme:1985jy}.  To illustrate
this, let us assume that \abc\ $\beta$-functions have \abc\ form
\beq
\begin{split}
\beta_{a} &=\frac{1}{16 \pi^2}[ \sum_{b,c,d\neq
  g}\beta^{(1)\,bcd}_{a}g_b g_c g_d+
\sum_{b\neq g}\beta^{(1)\,b}_{a}g_b g^2]+\cdots~,\\
\beta_{g} &=\frac{1}{16 \pi^2}\beta^{(1)}_{g}g^3+ \cdots~,
\end{split}
\eeq
where
$\cdots$ stands for higher order terms, \xyz\ $ \beta^{(1)\,bcd}_{a}$'s
are symmetric in $ b,c,d$.  We then assume that \abc\ $\rho_{a}^{(n)}$'s
with $n\leq r$ have been uniquely determined. To obtain
$\rho_{a}^{(r+1)}$'s, we insert \abc\ power series (\ref{powerser}) into
the REs (\ref{redeq}) \xyz\ collect terms of ${\cal O}(g^{2r+3})$ and
find
\beq
\sum_{d\neq g}M(r)_{a}^{d}\,\rho_{d}^{(r+1)} = \mbox{lower
  order quantities}~,\non
\eeq
where \abc\ r.h.s. is known by assumption,
and
\begin{align}
M(r)_{a}^{d} &=3\sum_{b,c\neq
  g}\,\beta^{(1)\,bcd}_{a}\,\rho_{b}^{(1)}\,
\rho_{c}^{(1)}+\beta^{(1)\,d}_{a}
-(2r+1)\,\beta^{(1)}_{g}\,\delta_{a}^{d}~,\label{M}\\
0 &=\sum_{b,c,d\neq g}\,\beta^{(1)\,bcd}_{a}\,
\rho_{b}^{(1)}\,\rho_{c}^{(1)}\,\rho_{d}^{(1)} +\sum_{d\neq
  g}\beta^{(1)\,d}_{a}\,\rho_{d}^{(1)}
-\beta^{(1)}_{g}\,\rho_{a}^{(1)}~,
\end{align}

 Therefore, \abc\ $\rho_{a}^{(n)}$'s for all $n > 1$ for a
given set of $\rho_{a}^{(1)}$'s can be uniquely determined if $\det
M(n)_{a}^{d} \neq 0$ for all $n \geq 0$.

As it will be clear later by examining specific examples, \abc\ various
couplings in SUSY theories have \abc\ same asymptotic
behaviour.  Therefore, searching for a power series solution of the
form (\ref{powerser}) to \abc\ REs (\ref{redeq}) is justified. This is
not \abc\ case in non-SUSY theories, although \abc\ deeper
reason for this fact is not fully understood.

The possibility of coupling unification described in this section
is without any doubt
attractive, because \abc\ ``completely reduced'' theory contains
only one independent coupling, but  it can be
unrealistic. Therefore, one often would like to impose fewer RGI
constraints, \xyz\ this is \abc\ idea of partial reduction \cite{Kubo:1985up,Kubo:1988zu}.

\subsection{Reduction of dimension one and two parameters}\label{sec:dimful}
The reduction of couplings was originally formulated for massless theories on the basis of
the Callan-Symanzik equation
\cite{Zimmermann:1984sx,Oehme:1984yy}.
The extension to theories with massive parameters is not
straightforward if one wants to keep the generality and the rigor on the same level as for the
massless case; one has to fulfill a set of requirements coming from the renormalization group
equations, the Callan-Symanzik equations etc. along with the normalization conditions
imposed on irreducible Green's functions
\cite{piguet1}.
There has been a lot of progress in this direction starting from ref. \cite{Kubo:1996js},
 where it was assumed that a mass-independent
renormalization scheme could be employed so that all the RG functions have only trivial
dependencies on dimensional parameters and then the mass parameters were
introduced similarly to couplings (i.e.\ as a power series in the
couplings). This choice was
justified later in \cite{zim2,Zimmermann:2001pq} where
the scheme independence of the reduction principle has been proven generally, i.e\ it was shown that apart from dimensionless couplings,
pole masses and gauge parameters, the model may also involve coupling parameters carrying a dimension and masses.
Therefore here, to simplify the analysis, we follow \citere{Kubo:1996js} and
we, too, use a mass-independent renormalization scheme.

We start by considering a renormalizable theory which contain a set of $(N + 1)$
dimension zero couplings, $\left(\hat g_0,\hat g_1, ...,\hat g_N\right)$,
a set of $L$ parameters with mass-dimension one, $\left(\hat h_1,...,\hat h_L\right)$,
and a set of $M$ parameters with mass-dimension two, $\left(\hat m_1^2,...,\hat m_M^2\right)$.
The renormalized irreducible vertex function $\Gamma$ satisfies the RG
equation
\beq
\label{RGE_OR_1}
\mathcal{D}\Gamma\left[\Phi's;\hat g_0,\hat g_1, ...,\hat g_N;\hat h_1,...,\hat h_L;\hat m_1^2,...,\hat m_M^2;\mu\right]=0~,
\eeq
where
\beq
\label{RGE_OR_2}
\mathcal{D}=\mu\frac{\partial}{\partial \mu}+
\sum_{i=0}^N \beta_i\frac{\partial}{\partial \hat g_i}+
\sum_{a=1}^L \gamma_a^h\frac{\partial}{\partial \hat h_a}+
\sum_{\alpha=1}^M \gamma_\alpha^{m^2}\frac{\partial}{\partial \hat m_\alpha ^2}+
\sum_J \Phi_I\gamma^{\phi I}_{\,\,\,\, J}\,\frac{\delta}{\delta\Phi_J}~,
\eeq
where $\mu$ is the energy scale,
while $\beta_i$ are the $\beta$-functions of the various dimensionless
couplings $g_i$, $\Phi_I$  are
the various matter fields and
$\ga_\alpha^{m^2}$, $\ga_a^h$ and $\ga^{\phi I}_{\,\,\,\, J}$
are the mass, trilinear coupling and wave function anomalous dimensions,
respectively
(where $I$ enumerates the matter fields).
In a mass independent renormalization scheme, the $\gamma$'s are given by
\beq
\label{gammas}
\begin{split}
\gamma^h_a&=\sum_{b=1}^L\gamma_a^{h,b}(g_0,g_1,...,g_N)\hat h_b,\\
\gamma_\alpha^{m^2}&=\sum_{\beta=1}^M \gamma_\alpha^{m^2,\beta}(g_0,g_1,...,g_N)\hat m_\beta^2+
\sum_{a,b=1}^L \gamma_\alpha^{m^2,ab}(g_0,g_1,...,g_N)\hat h_a\hat h_b,
\end{split}
\eeq
where $\gamma_a^{h,b}$, $\gamma_\alpha^{m^2,\beta}$ and $\gamma_\alpha^{m^2,ab}$ are power series of the
$g$'s (which are dimensionless) in perturbation theory.\\

\noindent We look for a reduced theory where
\[
g\equiv g_0,\qquad h_a\equiv \hat h_a\quad \textrm{for $1\leq a\leq P$},\qquad
m^2_\alpha\equiv\hat m^2_\alpha\quad \textrm{for $1\leq \alpha\leq Q$}
\]
are independent parameters and the reduction of the parameters left
\beq
\label{reduction}
\begin{split}
\hat g_i &= \hat g_i(g), \qquad (i=1,...,N),\\
\hat h_a &= \sum_{b=1}^P f_a^b(g)h_b, \qquad (a=P+1,...,L),\\
\hat m^2_\alpha &= \sum_{\beta=1}^Q e^\beta_\alpha(g)m^2_\beta + \sum_{a,b=1}^P k^{ab}_\alpha(g)h_ah_b,
\qquad (\alpha=Q+1,...,M)
\end{split}
\eeq
is consistent with the RG equations (\ref{RGE_OR_1},\ref{RGE_OR_2}). It
turns out that the following relations should be satisfied
\beq
\label{relation}
\begin{split}
\beta_g\,\frac{\partial\hat g_i}{\partial g} &= \beta_i,\qquad (i=1,...,N),\\
\beta_g\,\frac{\partial \hat h_a}{\partial g}+\sum_{b=1}^P \gamma^h_b\,\frac{\partial\hat h_a}{\partial h_b} &= \gamma^h_a,\qquad (a=P+1,...,L),\\
\beta_g\,\frac{\partial\hat m^2_\alpha}{\partial g} +\sum_{a=1}^P \gamma_a^h\,\frac{\partial\hat m^2_\alpha}{\partial h_a} +  \sum_{\beta=1}^Q \gamma_\beta ^{m^2}\,\frac{\partial\hat m_\alpha^2}{\partial m_\beta^2} &= \gamma_\alpha^{m^2}, \qquad (\alpha=Q+1,...,M).
\end{split}
\eeq
Using \refeqs{gammas} and (\ref{reduction}), the above relations reduce to
\beq
\label{relation_2}
\begin{split}
&\beta_g\,\frac{df^b_a}{dg}+ \sum_{c=1}^P f^c_a\left[\gamma^{h,b}_c + \sum_{d=P+1}^L \gamma^{h,d}_c f^b_d\right] -\gamma^{h,b}_a - \sum_{d=P+1}^L \gamma^{h,d}_a f^b_d=0,\\
&(a=P+1,...,L;\, b=1,...,P),\\
&\beta_g\,\frac{de^\beta_\alpha}{dg} + \sum_{\gamma=1}^Q e^\gamma_\alpha\left[\gamma_\gamma^{m^2,\beta} +
\sum _{\delta=Q+1}^M\gamma_\gamma^{m^2,\delta} e^\beta_\delta\right]-\gamma_\alpha^{m^2,\beta} -
\sum_{\delta=Q+1}^M \gamma_\alpha^{m^2,d}e^\beta_\delta =0,\\
&(\alpha=Q+1,...,Mq\, \beta=1,...,Q),\\
&\beta_g\,\frac{dk_\alpha^{ab}}{dg}
+ 2\sum_{c=1}^P \left(\gamma_c^{h,a} + \sum_{d=P+1}^L \gamma_c^{h,d} f_d^a\right)k_\alpha^{cb}
+\sum_{\beta=1}^Q e^\beta_\alpha\left[\gamma_\beta^{m^2,ab} + \sum_{c,d=P+1}^L \gamma_\beta^{m^2,cd}f^a_cf^b_d \right.\\
&\left. +2\sum_{c=P+1}^L \gamma_\beta^{m^2,cb}f^a_c + \sum_{\delta=Q+1}^M \gamma_\beta^{m^2,d} k_\delta^{ab}\right]- \left[\gamma_\alpha^{m^2,ab}+\sum_{c,d=P+1}^L \gamma_\alpha^{m^2,cd}f^a_c f^b_d\right.\\
&\left. +2 \sum_{c=P+1}^L \gamma_\alpha^{m^2,cb}f^a_c + \sum_{\delta=Q+1}^M \gamma_\alpha^{m^2,\delta}k_\delta^{ab}\right]=0,\\
&(\alpha=Q+1,...,M;\, a,b=1,...,P)~.
\end{split}
\eeq
The above relations ensure that the irreducible vertex function of the reduced theory
\beq
\label{Green}
\begin{split}
\Gamma_R&\left[\Phi\textrm{'s};g;h_1,...,h_P; m_1^2,...,m_Q^2;\mu\right]\equiv\\
&\Gamma \left[  \Phi\textrm{'s};g,\hat g_1(g)...,\hat g_N(g);
h_1,...,h_P,\hat h_{P+1}(g,h),...,\hat h_L(g,h);\right.\\
& \left.  \qquad\qquad\qquad m_1^2,...,m^2_Q,\hat m^2_{Q+1}(g,h,m^2),...,\hat m^2_M(g,h,m^2);\mu\right]
\end{split}
\eeq
has the same renormalization group flow as the original one.

The assumptions that the reduced theory is perturbatively renormalizable
means that the functions
$\hat g_i$, $f^b_a$, $e^\beta_\alpha$ and $k_\alpha^{ab}$, defined in (\ref{reduction}), should be
expressed as a power series in the primary coupling $g$:
\beq
\label{pert}
\begin{split}
\hat g_i & = g\sum_{n=0}^\infty \rho_i^{(n)} g^n,\qquad
f_a^b  =  g \sum_{n=0}^\infty \eta_a^{b(n)} g^n\\
e^\beta_\alpha & = \sum_{n=0}^\infty \xi^{\beta(n)}_\alpha g^n,\qquad
k_\alpha^{ab}=\sum_{n=0}^\infty \chi_\alpha^{ab(n)} g^n.
\end{split}
\eeq
The above expansion coefficients can be found by inserting these power
series into \refeqs{relation}, (\ref{relation_2}) and requiring the equations to be satisfied at each order of $g$.
It should be noted that the existence of a unique power series solution is a non-trivial matter: It depends on the theory
as well as on the choice of the set of independent parameters.

It should also be noted that in the case that there are no independent
mass-dimension one parameters ($\hat h$) the reduction of these terms take
naturally the form
\[
\hat h_a = \sum_{b=1}^L f_a^b(g)M,
\]
where $M$ is a mass-dimension one parameter which could be a gaugino mass which corresponds to the independent (gauge) coupling. In case, on top of that, there are no independent
mass-dimension two parameters ($\hat m^2$), the corresponding reduction takes analogous form
\[
\hat m^2_a=\sum_{b=1}^M e_a^b(g) M^2.
\]

\section{Finiteness in N = 1 Supersymmetric Gauge Theories}
Let us consider a chiral, anomaly free,
$N=1$ globally SUSY
gauge theory based on a group G with gauge coupling
constant $g$. The
superpotential of \abc\ theory is given by
\beq
W= \frac{1}{2}\,m_{ij} \,\phi_{i}\,\phi_{j}+
\frac{1}{6}\,C_{ijk} \,\phi_{i}\,\phi_{j}\,\phi_{k}~,
\label{supot}
\eeq
where $m_{ij}$ \xyz\ $C_{ijk}$ are gauge invariant tensors and
the matter field $\phi_{i}$ transforms
according to \abc\ irreducible representation  $R_{i}$
of \abc\ gauge group $G$. The
renormalization constants associated with the
superpotential (\ref{supot}), assuming that
SUSY is preserved, are
\begin{align}
\phi_{i}^{0}&=(Z^{j}_{i})^{(1/2)}\,\phi_{j}~,~\\
m_{ij}^{0}&=Z^{i'j'}_{ij}\,m_{i'j'}~,~\\
C_{ijk}^{0}&=Z^{i'j'k'}_{ijk}\,C_{i'j'k'}~.
\end{align}
The $N=1$ non-renormalization theorem \cite{Wess:1973kz,Iliopoulos:1974zv,Fujikawa:1974ay} ensures that
there are no mass
and cubic-interaction-term infinities \xyz\ therefore
\beq
\begin{split}
Z_{ijk}^{i'j'k'}\,Z^{1/2\,i''}_{i'}\,Z^{1/2\,j''}_{j'}
\,Z^{1/2\,k''}_{k'}&=\delta_{(i}^{i''}
\,\delta_{j}^{j''}\delta_{k)}^{k''}~,\\
Z_{ij}^{i'j'}\,Z^{1/2\,i''}_{i'}\,Z^{1/2\,j''}_{j'}
&=\delta_{(i}^{i''}
\,\delta_{j)}^{j''}~.
\end{split}
\eeq
As a result \abc\ only surviving possible infinities are
the wave-function renormalization constants
$Z^{j}_{i}$, i.e.  one infinity
for each field. \Abc\ one-loop $\beta$-function of \abc\ gauge
coupling $g$ is given by \cite{Parkes:1984dh}
\beq
\beta^{(1)}_{g}=\frac{d g}{d t} =
\frac{g^3}{16\pi^2}\,[\,\sum_{i}\,l(R_{i})-3\,C_{2}(G)\,]~,
\label{betag}
\eeq
where $l(R_{i})$ is \abc\ Dynkin index of $R_{i}$ \xyz\ $C_{2}(G)$
 is the
quadratic Casimir invariant of \abc\ adjoint representation of the
gauge group $G$. \Abc\ $\beta$-functions of
$C_{ijk}$,
by virtue of \abc\ non-renormalization theorem, are related to the
anomalous dimension matrix $\gamma_{ij}$ of \abc\ matter fields
$\phi_{i}$ as:
\beq
\beta_{ijk} =
 \frac{d C_{ijk}}{d t}~=~C_{ijl}\,\gamma^{l}_{k}+
 C_{ikl}\,\gamma^{l}_{j}+
 C_{jkl}\,\gamma^{l}_{i}~.
\label{betay}
\eeq
At one-loop level $\gamma_{ij}$ is \cite{Parkes:1984dh}
\beq
\gamma^{i(1)}_j=\frac{1}{32\pi^2}\,[\,
C^{ikl}\,C_{jkl}-2\,g^2\,C_{2}(R)\delta_{j}^i\,],
\label{gamay}
\eeq
where $C_{2}(R)$ is \abc\ quadratic Casimir invariant of \abc\ representation
$R_{i}$, \xyz\ $C^{ijk}=C_{ijk}^{*}$.
%\marginpar{\htb{corrected here \abc\ formula \xyz\ \abc\ indices}}
%
Since
dimensional coupling parameters such as masses  \xyz\ couplings of cubic
scalar field terms do not influence \abc\ asymptotic properties
 of a theory on which we are interested here, it is
sufficient to take into account only \abc\ dimensionless SUSY
couplings such as $g$ \xyz\ $C_{ijk}$.
So we neglect \abc\ existence of dimensional parameters and
assume furthermore that
$C_{ijk}$ are real so that $C_{ijk}^2$ are always positive numbers.

As one can see from Eqs.~(\ref{betag}) \xyz\ (\ref{gamay}),
 all \abc\ one-loop $\beta$-functions of \abc\ theory vanish if
 $\beta_g^{(1)}$ \xyz\ $\gamma _{ij}^{(1)}$ vanish, i.e.
\begin{equation}
\sum _i \ell (R_i) = 3 C_2(G) \,,
\label{1st}
\end{equation}

\begin{equation}
C^{ikl} C_{jkl} = 2\delta ^i_j g^2  C_2(R_i)\,,
\label{2nd}
\end{equation}

The conditions for finiteness for $N=1$ field theories with $SU(N)$ gauge
symmetry are discussed in \cite{Rajpoot:1984zq}, \xyz\ the
analysis of \abc\ anomaly-free \xyz\ no-charge renormalization
requirements for these theories can be found in \cite{Rajpoot:1985aq}.
A very interesting result is that \abc\ conditions (\ref{1st},\ref{2nd})
are necessary \xyz\ sufficient for finiteness at \abc\ two-loop level
\cite{Parkes:1984dh,West:1984dg,Jones:1985ay,Jones:1984cx,Parkes:1985hh}.

In case SUSY is broken by soft terms, \abc\ requirement of
finiteness in \abc\ one-loop soft breaking terms imposes further
constraints among themselves \cite{Jones:1984cu}.  In addition, \abc\ same set
of conditions that are sufficient for one-loop finiteness of \abc\ soft
breaking terms renders \abc\ soft sector of \abc\ theory two-loop
finite\cite{Jack:1994kd}.

The one- \xyz\ two-loop finiteness conditions (\ref{1st},\ref{2nd}) restrict
considerably \abc\ possible choices of \abc\ irreducible representations
(irreps)
$R_i$ for a given
group $G$ as well as \abc\ Yukawa couplings in \abc\ superpotential
(\ref{supot}).  Note in particular that \abc\ finiteness conditions cannot be
applied to \abc\ minimal SUSY standard model (MSSM), since \abc\ presence
of a $U(1)$ gauge group is incompatible with \abc\ condition
(\ref{1st}), due to $C_2[U(1)]=0$.  This naturally leads to the
expectation that finiteness should be attained at \abc\ grand unified
level only, \abc\ MSSM being just \abc\ corresponding, low-energy,
effective theory.

Another important consequence of one- \xyz\ two-loop finiteness is that
SUSY (most probably) can only be broken due to \abc\ soft
breaking terms.  Indeed, due to \abc\ unacceptability of gauge singlets,
F-type spontaneous symmetry breaking \cite{O'Raifeartaigh:1975pr}
terms are incompatible with finiteness, as well as D-type
\cite{Fayet:1974jb} spontaneous breaking which requires \abc\ existence
of a $U(1)$ gauge group.

A natural question to ask is what happens at higher-loop orders.  The
answer is contained in a theorem
\cite{Lucchesi:1987he,Lucchesi:1987ef} which states \abc\ necessary and
sufficient conditions to achieve finiteness at all orders.  Before we
discuss \abc\ theorem let us make some introductory remarks.  The
finiteness conditions impose relations between gauge \xyz\ Yukawa
couplings.  To require such relations which render \abc\ couplings
mutually dependent at a given renormalization point is trivial.  What
is not trivial is to guarantee that relations leading to a reduction
of \abc\ couplings hold at any renormalization point.  As we have seen,
the necessary \xyz\ also sufficient condition for this to happen is to
require that such relations are solutions to \abc\ REs
\beq \beta _g
\frac{d C_{ijk}}{dg} = \beta _{ijk}
\label{redeq2}
\eeq
and hold at all orders.   Remarkably, \abc\ existence of
all-order power series solutions to (\ref{redeq2}) can be decided at
one-loop level, as already mentioned.

Let us now turn to \abc\ all-order finiteness theorem
\cite{Lucchesi:1987he,Lucchesi:1987ef}, which states the conditions under which an $N=1$
SUSY gauge theory can become finite to all orders in the
sense of vanishing $\beta$-functions, that is of physical scale
invariance.  It is based on (a) \abc\ structure of \abc\ supercurrent in
$N=1$ SUSY gauge theory
\cite{Ferrara:1974pz,Piguet:1981mu,Piguet:1981mw}, \xyz\ on (b) the
non-renormalization properties of $N=1$ chiral anomalies
\cite{Lucchesi:1987he,Lucchesi:1987ef,Piguet:1986td,Piguet:1986pk,Ensign:1987wy}.
Details on \abc\ proof can be found in
refs. \cite{Lucchesi:1987he,Lucchesi:1987ef} \xyz\ further discussion in
\citeres{Piguet:1986td,Piguet:1986pk,Ensign:1987wy,Lucchesi:1996ir,Piguet:1996mx}.
Here, following mostly \citere{Piguet:1996mx} we present a
comprehensible sketch of \abc\ proof.

Consider an $N=1$ SUSY gauge theory, with simple Lie group
$G$.  \Abc\ content of this theory is given at \abc\ classical level by
the matter supermultiplets $S_i$, which contain a scalar field
$\phi_i$ \xyz\ a Weyl spinor $\psi_{ia}$, \xyz\ \abc\ vector supermultiplet
$V_a$, which contains a gauge vector field $A_{\mu}^a$ \xyz\ a gaugino
Weyl spinor $\lambda^a_{\alpha}$.

Let us first recall certain facts about \abc\ theory:

\noindent (1)  A massless $N=1$ SUSY theory is invariant
under a $U(1)$ chiral transformation $R$ under which \abc\ various fields
transform as follows
\beq
\begin{split}
A'_{\mu}&=A_{\mu},~~\lambda '_{\alpha}=\exp({-i\theta})\lambda_{\alpha}\\
\phi '&= \exp({-i\frac{2}{3}\theta})\phi,~~\psi_{\alpha}'= \exp({-i\frac{1}
    {3}\theta})\psi_{\alpha},~\cdots
\end{split}
\eeq
The corresponding axial Noether current $J^{\mu}_R(x)$ is
\beq
J^{\mu}_R(x)=\bar{\lambda}\gamma^{\mu}\gamma^5\lambda + \cdots
\label{noethcurr}
\eeq
is conserved classically, while in \abc\ quantum case is violated by the
axial anomaly
\beq
\partial_{\mu} J^{\mu}_R =
r(\epsilon^{\mu\nu\sigma\rho}F_{\mu\nu}F_{\sigma\rho}+\cdots).
\label{anomaly}
\eeq

From its known topological origin in ordinary gauge theories
\cite{AlvarezGaume:1983cs,Bardeen:1984pm,Zumino:1983rz}, one would
expect \abc\ axial vector current
$J^{\mu}_R$ to satisfy \abc\ Adler-Bardeen theorem  and
receive corrections only at \abc\ one-loop level.  Indeed it has been
shown that \abc\ same non-renormalization theorem holds also in
SUSY theories \cite{Piguet:1986td,Piguet:1986pk,Ensign:1987wy}.  Therefore
\beq
r=\hbar \beta_g^{(1)}.
\label{r}
\eeq

\noindent (2)  \Abc\ massless theory we consider is scale invariant at
the classical level and, in general, there is a scale anomaly due to
radiative corrections.  \Abc\ scale anomaly appears in \abc\ trace of the
energy momentum tensor $T_{\mu\nu}$, which is traceless classically.
It has \abc\ form
\beq
T^{\mu}_{\mu} = \beta_g F^{\mu\nu}F_{\mu\nu} +\cdots
\label{Tmm}
\eeq

\noindent (3)  Massless, $N=1$ SUSY gauge theories are
classically invariant under \abc\ SUSY extension of the
conformal group -- \abc\ superconformal group.  Examining the
superconformal algebra, it can be seen that \abc\ subset of
superconformal transformations consisting of translations,
SUSY transformations, \xyz\ axial $R$ transformations is closed
under SUSY, i.e. these transformations form a representation
of SUSY.  It follows that \abc\ conserved currents
corresponding to these transformations make up a supermultiplet
represented by an axial vector superfield called \abc\ supercurrent~$J$,
\beq
J \equiv \{ J'^{\mu}_R, ~Q^{\mu}_{\alpha}, ~T^{\mu}_{\nu} , ... \},
\label{J}
\eeq
where $J'^{\mu}_R$ is \abc\ current associated to R invariance,
$Q^{\mu}_{\alpha}$ is \abc\ one associated to SUSY invariance,
and $T^{\mu}_{\nu}$ \abc\ one associated to translational invariance
(energy-momentum tensor).

The anomalies of \abc\ R current $J'^{\mu}_R$, \abc\ trace
anomalies of \abc\
SUSY current, \xyz\ \abc\ energy-momentum tensor, form also
a second supermultiplet, called \abc\ supertrace anomaly
\beq
\begin{split}
S &= \{ Re~ S, ~Im~ S,~S_{\alpha}\} =\non\\
& \{T^{\mu}_{\mu},~\partial _{\mu} J'^{\mu}_R,~\sigma^{\mu}_{\alpha
  \dot{\beta}} \bar{Q}^{\dot\beta}_{\mu}~+~\cdots \}
\end{split}
\eeq
where $T^{\mu}_{\mu}$ is given in Eq.(\ref{Tmm}) and
\begin{align}
\partial _{\mu} J'^{\mu}_R &~=~\beta_g\epsilon^{\mu\nu\sigma\rho}
F_{\mu\nu}F_{\sigma\rho}+\cdots\\
\sigma^{\mu}_{\alpha \dot{\beta}} \bar{Q}^{\dot\beta}_{\mu}&~=~\beta_g
\lambda^{\beta}\sigma^{\mu\nu}_{\alpha\beta}F_{\mu\nu}+\cdots
\end{align}

\noindent (4) It is very important to note that
the Noether current defined in (\ref{noethcurr}) is not \abc\ same as the
current associated to R invariance that appears in the
supercurrent
$J$ in (\ref{J}), but they coincide in \abc\ tree approximation.
So starting from a unique classical Noether current
$J^{\mu}_{R(class)}$,  \abc\ Noether
current $J^{\mu}_R$ is defined as \abc\ quantum extension of
$J^{\mu}_{R(class)}$ which allows for the
validity of \abc\ non-renormalization theorem.  On \abc\ other hand
$J'^{\mu}_R$, is defined to belong to \abc\ supercurrent $J$,
together with \abc\ energy-momentum tensor.  \Abc\ two requirements
cannot be fulfilled by a single current operator at \abc\ same time.

Although \abc\ Noether current $J^{\mu}_R$ which obeys (\ref{anomaly})
and \abc\ current $J'^{\mu}_R$ belonging to \abc\ supercurrent multiplet
$J$ are not \abc\ same, there is a relation
\cite{Lucchesi:1987he,Lucchesi:1987ef} between quantities associated
with them
\beq
r=\beta_g(1+x_g)+\beta_{ijk}x^{ijk}-\gamma_Ar^A
\label{rbeta}
\eeq
where $r$ was given in Eq.(\ref{r}).  \Abc\ $r^A$ are the
non-renormalized coefficients of
the anomalies of \abc\ Noether currents associated to \abc\ chiral
invariances of \abc\ superpotential, \xyz\ --like $r$-- are strictly
one-loop quantities. \Abc\ $\gamma_A$'s are linear
combinations of \abc\ anomalous dimensions of \abc\ matter fields, and
$x_g$, \xyz\ $x^{ijk}$ are radiative correction quantities.
The structure of equality (\ref{rbeta}) is independent of the
renormalization scheme.

One-loop finiteness, i.e. vanishing of \abc\ $\beta$-functions at one loop,
implies that \abc\ Yukawa couplings $C_{ijk}$ must be functions of
the gauge coupling $g$. To find a similar condition to all orders it
is necessary \xyz\ sufficient for \abc\ Yukawa couplings to be a formal
power series in $g$, which is solution of \abc\ REs (\ref{redeq2}).

We can now state \abc\ theorem for all-order vanishing
$\beta$-functions.
\bigskip

{\bf Theorem:}

Consider an $N=1$ SUSY Yang-Mills theory, with simple gauge
group. If \abc\ following conditions are satisfied:
\begin{enumerate}
\item There is no gauge anomaly.
\item \Abc\ gauge $\beta$-function vanishes at one loop
  \beq
  \beta^{(1)}_g = 0 =\sum_i l(R_{i})-3\,C_{2}(G).
  \eeq
\item There exist solutions of \abc\ form
  \beq
  C_{ijk}=\rho_{ijk}g,~\qquad \rho_{ijk}\in\complex
  \label{soltheo}
  \eeq
to \abc\  conditions of vanishing one-loop matter fields anomalous
dimensions
%\marginpar{\htb{corrected $C_2(R)$}}
\beq
\begin{split}
  &\gamma^{i~(1)}_j~=~0\\
  &=\frac{1}{32\pi^2}~[ ~
  C^{ikl}\,C_{jkl}-2~g^2~C_{2}(R)\delta_j^i ].
\end{split}
\eeq
\item These solutions are isolated \xyz\ non-degenerate when considered
  as solutions of vanishing one-loop Yukawa $\beta$-functions:
   \beq
   \beta_{ijk}=0.
   \eeq
\end{enumerate}
Then, each of \abc\ solutions (\ref{soltheo}) can be uniquely extended
to a formal power series in $g$, \xyz\ \abc\ associated super Yang-Mills
models depend on \abc\ single coupling constant $g$ with a $\beta$
function which vanishes at all orders.

\bigskip

It is important to note a few things:
The requirement of isolated \xyz\ non-degenerate
solutions guarantees \abc\
existence of a unique formal power series solution to \abc\ reduction
equations.
The vanishing of \abc\ gauge $\beta$~function at one loop,
$\beta_g^{(1)}$, is equivalent to \abc\
vanishing of \abc\ R current anomaly (\ref{anomaly}).  \Abc\ vanishing of
the anomalous
dimensions at one loop implies \abc\ vanishing of \abc\ Yukawa couplings
$\beta$~functions at that order.  It also implies \abc\ vanishing of the
chiral anomaly coefficients $r^A$.  This last property is a necessary
condition for having $\beta$ functions vanishing at all orders.\footnote{There is an alternative way to find finite theories
  \cite{Leigh:1995ep}.}

\bigskip

{\bf Proof:}

Insert $\beta_{ijk}$ as given by \abc\ REs into the
relationship (\ref{rbeta}) between \abc\ axial anomalies coefficients and
the $\beta$-functions.  Since these chiral anomalies vanish, we get
for $\beta_g$ an homogeneous equation of \abc\ form
\beq
0=\beta_g(1+O(\hbar)).
\label{prooftheo}
\eeq
The solution of this equation in \abc\ sense of a formal power series in
$\hbar$ is $\beta_g=0$, order by order.  Therefore, due to the
REs (\ref{redeq2}), $\beta_{ijk}=0$ too.

Thus we see that finiteness \xyz\ reduction of couplings are intimately
related. Since an equation like Eq.(\ref{rbeta}) is lacking in
non-SUSY theories, one cannot extend \abc\ validity of a
similar theorem in such theories.

\section{The SSB sector of reduced $N=1$ SUSY and Finite Theories}

As we have seen in subsection \ref{sec:dimful},
the method of reducing \abc\ dimensionless couplings has been
extended\cite{Kubo:1996js}, to \abc\ soft SUSY breaking (SSB)
dimensionful parameters of $N = 1$ SUSY theories.  In
addition it was found \cite{Kawamura:1997cw} that RGI SSB scalar
masses in GYU  models satisfy a universal sum rule.
%Here we will describe first how \abc\ use of \abc\ available two-loop RG
%functions \xyz\ \abc\ requirement of finiteness of \abc\ SSB parameters up
%to this order leads to \abc\ soft scalar-mass sum rule
%\cite{Kobayashi:1997qx}.

Consider \abc\ superpotential given by (\ref{supot})
along with \abc\ Lagrangian for SSB terms
\beq
\begin{split}
-{\cal L}_{\rm SSB} &=
\frac{1}{6} \,h^{ijk}\,\phi_i \phi_j \phi_k
+
\frac{1}{2} \,b^{ij}\,\phi_i \phi_j \non\\
&+
\frac{1}{2} \,(m^2)^{j}_{i}\,\phi^{*\,i} \phi_j+
\frac{1}{2} \,M\,\lambda \lambda+\mbox{h.c.},
\end{split}
\eeq
where \abc\ $\phi_i$ are the
scalar parts of \abc\ chiral superfields $\Phi_i$ , $\lambda$ are \abc\ gauginos
and $M$ their unified mass.

We assume that \abc\ reduction equations
admit power series solutions of \abc\ form
\beq
C^{ijk} = g\,\sum_{n}\,\rho^{ijk}_{(n)} g^{2n}~.
\label{Yg}
\eeq

If we knew higher-loop $\beta$-functions explicitly, we could follow \abc\ same
procedure \xyz\ find higher-loop RGI relations among SSB terms.
However, \abc\ $\beta$-functions of \abc\ soft scalar masses are explicitly
known only up to two loops.
In order to obtain higher-loop results some relations among
$\beta$-functions are needed.

In the case of finite theories we assume that
the gauge group is a simple group \xyz\ \abc\ one-loop
$\beta$-function of \abc\
gauge coupling $g$  vanishes.
According to \abc\ finiteness theorem
of \citeres{Lucchesi:1987ef,Lucchesi:1987he} and the assumption given in (\ref{Yg}), \abc\ theory is then finite to all orders in
perturbation theory, if, among others, \abc\ one-loop anomalous dimensions
$\gamma_{i}^{j(1)}$ vanish.  \Abc\ one- \xyz\ two-loop finiteness for
$h^{ijk}$ can be achieved by \cite{Jack:1994kd}
\beq
h^{ijk} = -M C^{ijk}+\dots =-M
\rho^{ijk}_{(0)}\,g+O(g^5)~,
\label{hY}
\eeq
where $\dots$ stand for  higher order terms.

Now, to obtain \abc\ two-loop sum rule for
soft scalar masses, we assume that
the lowest order coefficients $\rho^{ijk}_{(0)}$
and also $(m^2)^{i}_{j}$ satisfy \abc\ diagonality relations
\beq
\rho_{ipq(0)}\rho^{jpq}_{(0)} \propto  \delta_{i}^{j}~\mbox{for all}
~p ~\mbox{and}~q~~\mbox{and}~~
(m^2)^{i}_{j}= m^{2}_{j}\delta^{i}_{j}~,
\label{cond1}
\eeq
respectively.
Then we find \abc\ following soft scalar-mass sum
rule \cite{Kobayashi:1997qx,Kobayashi:1999pn,Mondragon:2003bp}
\beq
(~m_{i}^{2}+m_{j}^{2}+m_{k}^{2}~)/
M M^{\dag} =
1+\frac{g^2}{16 \pi^2}\,\Delta^{(2)}+O(g^4)~
\label{sumr}
\eeq
for i, j, k with $\rho^{ijk}_{(0)} \neq 0$, where $\Delta^{(2)}$ is
the two-loop correction
\beq
\Delta^{(2)} =  -2\sum_{l} [(m^{2}_{l}/M M^{\dag})-(1/3)]~T(R_l),
\label{delta}
\eeq
which vanishes for the
universal choice in accordance with \abc\ previous findings of
\citere{Jack:1994kd} (in the above relation $T(R_l)$ is \abc\ Dynkin index of the
$R_l$ irrep).
%\marginpar{\htb{T was not defined, since we changed notation from
%    $l(R_i)$ to $T(R_i)$}}

Making use of \abc\ spurion technique
\cite{Delbourgo:1974jg,Salam:1974pp,Fujikawa:1974ay,Grisaru:1979wc,Girardello:1981wz}, it is possible to find
the following  all-loop relations among SSB $\beta$-functions,
\cite{Hisano:1997ua,Jack:1997pa,Avdeev:1997vx,Kazakov:1998uj,Kazakov:1997nf,Jack:1997eh}
\begin{align}
\beta_M &= 2{\cal O}\left(\frac{\beta_g}{g}\right)~,
\label{betaM}\\
\beta_h^{ijk}&=\gamma^i{}_lh^{ljk}+\gamma^j{}_lh^{ilk}
+\gamma^k{}_lh^{ijl}\non\\
&-2\gamma_1^i{}_lC^{ljk}
-2\gamma_1^j{}_lC^{ilk}-2\gamma_1^k{}_lC^{ijl}~,\label{betah}\\
(\beta_{m^2})^i{}_j &=\left[ \Delta
+ X \frac{\partial}{\partial g}\right]\gamma^i{}_j~,
\label{betam2}\\
{\cal O} &=\left(Mg^2\frac{\partial}{\partial g^2}
-h^{lmn}\frac{\partial}{\partial C^{lmn}}\right)~,
\label{diffo}\\
\Delta &= 2{\cal O}{\cal O}^* +2|M|^2 g^2\frac{\partial}
{\partial g^2} +\tilde{C}_{lmn}
\frac{\partial}{\partial C_{lmn}} +
\tilde{C}^{lmn}\frac{\partial}{\partial C^{lmn}}~,
\end{align}
where $(\gamma_1)^i{}_j={\cal O}\gamma^i{}_j$,
$C_{lmn} = (C^{lmn})^*$, \xyz\
\beq
\tilde{C}^{ijk}=
(m^2)^i{}_lC^{ljk}+(m^2)^j{}_lC^{ilk}+(m^2)^k{}_lC^{ijl}~.
\label{tildeC}
\eeq
Assuming, following \cite{Jack:1997pa},  that \abc\ relation
\beq
h^{ijk} = -M (C^{ijk})'
\equiv -M \frac{d C^{ijk}(g)}{d \ln g}~,
\label{h2}
\eeq
among couplings is all-loop RGI and using \abc\ all-loop gauge
$\beta$-function of Novikov {\em et al.}
\cite{Novikov:1983ee,Novikov:1985rd,Shifman:1996iy} given
by
\beq
\beta_g^{\rm NSVZ} =
\frac{g^3}{16\pi^2}
\left[ \frac{\sum_l T(R_l)(1-\gamma_l /2)
-3 C(G)}{ 1-g^2C(G)/8\pi^2}\right]~,
\label{bnsvz}
\eeq
\abc\ all-loop RGI sum rule \cite{Kobayashi:1998jq} was found:
\beq
\label{sum2}
\begin{split}
m^2_i+m^2_j+m^2_k &=
|M|^2 \{~
\frac{1}{1-g^2 C(G)/(8\pi^2)}\frac{d \ln C^{ijk}}{d \ln g}
+\frac{1}{2}\frac{d^2 \ln C^{ijk}}{d (\ln g)^2}~\}\\
& +\sum_l
\frac{m^2_l T(R_l)}{C(G)-8\pi^2/g^2}
\frac{d \ln C^{ijk}}{d \ln g}~.
\end{split}
\eeq
In addition
the exact-$\beta$-function for $m^2$
in \abc\ NSVZ scheme has been obtained \cite{Kobayashi:1998jq} for \abc\ first time and
is given by
\beq
\label{bm23}
\begin{split}
\beta_{m^2_i}^{\rm NSVZ} &=\left[~
|M|^2 \{~
\frac{1}{1-g^2 C(G)/(8\pi^2)}\frac{d }{d \ln g}
+\frac{1}{2}\frac{d^2 }{d (\ln g)^2}~\}\right. \\
& \left. +\sum_l
\frac{m^2_l T(R_l)}{C(G)-8\pi^2/g^2}
\frac{d }{d \ln g}~\right]~\gamma_{i}^{\rm NSVZ}~.
\end{split}
\eeq
Surprisingly enough, \abc\ all-loop result (\ref{sum2}) coincides with
the superstring result for \abc\ finite case in a certain class of
orbifold models \cite{Kobayashi:1997qx} if
$d \ln C^{ijk}/{d \ln g}=1$.

\subsection{All-loop RGI Relations in the SSB Sector}
Let us now see how the all-loop results on the SSB $\beta$-functions, Eqs. (\ref{betaM})-(\ref{tildeC}), lead to all-loop RGI relations. We make two assumptions:\\
(a) the existence of an RGI surface on which $C=C(g)$, or equivalently that
\beq
\frac{dC^{ijk}}{dg}=\frac{\beta_C^{ijk}}{\beta_g}\label{4.1.1}
\eeq
holds, i.e. reduction of couplings is possible, and\\
(b) the existence of an RGI surface on which
\beq
h^{ijk}=-M\frac{dC(g)^{ijk}}{d\ln{g}}
\eeq
holds, too, in all orders.

Then one can prove, \cite{Jack:1999aj,Kobayashi:1998iaa}, that the following relations are RGI to all loops (note that in both  (a) and (b) assumptions above we do not rely on specific solutions of these equations)
\begin{align}
M&=M_0\frac{\beta_g}{g}~,\label{4.1.3}\\
h^{ijk}&=-M_0\beta_C^{ijk}~,\\
b^{ij}&=-M_0\beta_{\mu}^{ij}~,\label{bij}\\
(m^2)^i_{\phantom{a}j}&=\frac{1}{2}|M_0|^2\mu\frac{d\gamma^i_{\phantom{a}j}}{d\mu}\label{m2ij}
\end{align}
where $M_0$ is an arbitrary reference mass scale to be specified shortly. The assumption that
\beq
C_{\alpha}\frac{\partial}{\partial C_{\alpha}}=C_{\alpha}^*\frac{\partial}{\partial C_{\alpha}}
\eeq
for an RGI surface $F(g,C^{ijk},C^{*ijk})$ leads to
\beq
\frac{d}{dg}=\Big(\frac{\partial}{\partial g}+2\frac{\partial}{\partial C}\frac{dC}{dg}\Big)=\Big(\frac{\partial}{\partial g}+2\frac{\beta_C}{\beta_g}\frac{\partial}{\partial C}\Big)~,
\eeq
where Eq.(\ref{4.1.1}) has been used. Now let us consider the partial differential operator $\cal O$ in Eq.(\ref{diffo}) which, assuming Eq.(\ref{h2}) becomes
\beq
{\cal O} =\frac{1}{2}M\frac{d}{d\ln{g}}~.
\eeq 
In turn, $\beta_M$ given in Eq.(\ref{betaM}) becomes
\beq
\beta_M=M\frac{d}{d\ln{g}}(\frac{\beta_g}{g})~,\label{4.1.10}
\eeq
which by integration provides us \cite{Jack:1999aj,Karch:1998qa} with the generalized, i.e. including Yukawa couplings, all-loop RGI Hisano - Shifman relation \cite{Hisano:1997ua}
\beq
M=\frac{\beta_g}{g}M_0~,\label{Hisano-Shifman}
\eeq
where $M_0$ is the integration constant and can be associated to the unification scale $M_U$ in GUTs or to the gravitino mass $m_{3/2}$ in a supergravity framework. Therefore, Eq.(\ref{Hisano-Shifman}) becomes the all-loop RGE Eq.(\ref{4.1.3}). Note that $\beta_M$ using Eqs. (\ref{4.1.10}) and (\ref{Hisano-Shifman}) can be written as
\beq
\beta_M=M_0\frac{d}{dt}(\beta_g/g)~.
\eeq
Similarly
\beq
(\gamma_1)^i_{\phantom{a}j}={\cal O}\gamma^i_{\phantom{a}j}=\frac{1}{2}M_0\frac{d\gamma^i_{\phantom{a}j}}{dt}~.\label{gamma1}
\eeq
Next, form Eq.(\ref{h2}) and (\ref{Hisano-Shifman}) we obtain
\beq
h^{ijk}=-M_0\beta_C^{ijk}~,\label{hijk}
\eeq
while $\beta_h^{ijk}$, given in Eq.(\ref{betah}) and using Eq.(\ref{gamma1}), becomes \cite{Jack:1999aj}
\beq
\beta_h^{ijk}=M_0\frac{d}{dt}\beta_C^{ijk}
\eeq
which shows that Eq.(\ref{hijk}) is all-loop RGI. In a similar way Eq.(\ref{bij}) can be shown to be all-loop RGI.

Finally, we would like to emphasize that under the same assumptions (a) and (b) the sum rule given in Eq.(\ref{sum2}) has been proven \cite{Kobayashi:1998jq} to be all-loop RGI, which (using Eq.(\ref{Hisano-Shifman})) gives us a generalization of Eq.(\ref{m2ij}) to be applied in considerations of non-universal soft scalar masses, which are necessary in many cases. Moreover, the sum rule holds also in the more general cases, discussed in
subsection 2.1, according to which exact relations among the squared
scalar and gaugino masses can be found.

\section{A successful Finite Unified Theory}

\renewcommand{\FUTB}{{\bf \boldmath{$SU(5)$}-FUT}}
\newcommand{\FUTBm}{{\bf \boldmath{$SU(5)$}-FUT} (with $\mu < 0$)}

%\section{The (so far) best Finite Unified Theory}

%\subsection{Definition of \FUTB}

We review an all-loop FUT
with $SU(5)$ as gauge group, where \abc\ reduction of couplings has been
applied to \abc\ third generation of quarks \xyz\ leptons. This FUT was selected previously on the basis of agreement with the known experimental data at the time \cite{Heinemeyer:2007tz} and was predicting the Higgs mass to be in the range $121-126$ GeV almost five years before the discovery.
The particle content of \abc\ model we will study, which we denote \FUTB\,
consists of \abc\
following supermultiplets: three ($\overline{\bf 5} + \bf{10}$),
needed for each of \abc\ three generations of quarks \xyz\ leptons, four
($\overline{\bf 5} + {\bf 5}$) \xyz\ one ${\bf 24}$ considered as Higgs
supermultiplets.
When \abc\ gauge group of \abc\ finite GUT is broken \abc\ theory is no
longer finite, \xyz\ we will assume that we are left with \abc\ MSSM
\cite{Kapetanakis:1992vx,Kubo:1994bj,Kubo:1994xa,Kubo:1995hm,Kubo:1997fi,Mondragon:1993tw}.

A predictive GYU  $SU(5)$ model which is finite to all
orders, in addition to \abc\ requirements mentioned already, should also
have \abc\ following properties:
\begin{enumerate}
\item
One-loop anomalous dimensions are diagonal,
i.e.,  $\gamma_{i}^{(1)\,j} \propto \delta^{j}_{i} $.
\item Three fermion generations in \abc\ irreducible representations
  $\overline{\bf 5}_{i},{\bf 10}_i~(i=1,2,3)$ which obviously should
  not couple to \abc\ adjoint ${\bf 24}$.
\item \Abc\ two Higgs doublets of \abc\ MSSM should mostly be made out of a
pair of Higgs quintet \xyz\ anti-quintet, which couple to \abc\ third
generation.
\end{enumerate}

After \abc\ reduction
of couplings \abc\ symmetry is enhanced, leading to \abc\
following superpotential \cite{Kobayashi:1997qx,Mondragon:2009zz}
\begin{align}
W &= \sum_{i=1}^{3}\,[~\frac{1}{2}g_{i}^{u}
\,{\bf 10}_i{\bf 10}_i H_{i}+
g_{i}^{d}\,{\bf 10}_i \overline{\bf 5}_{i}\,
\overline{H}_{i}~] +
g_{23}^{u}\,{\bf 10}_2{\bf 10}_3 H_{4} \\
 &+g_{23}^{d}\,{\bf 10}_2 \overline{\bf 5}_{3}\,
\overline{H}_{4}+
g_{32}^{d}\,{\bf 10}_3 \overline{\bf 5}_{2}\,
\overline{H}_{4}+
g_{2}^{f}\,H_{2}\,
{\bf 24}\,\overline{H}_{2}+ g_{3}^{f}\,H_{3}\,
{\bf 24}\,\overline{H}_{3}+
\frac{g^{\lambda}}{3}\,({\bf 24})^3~.\nonumber
\label{w-futb}
\end{align}
The non-degenerate \xyz\ isolated solutions to
$\gamma^{(1)}_{i}=0$ give us:
\BEA
&& (g_{1}^{u})^2
=\frac{8}{5}~ g^2~, ~(g_{1}^{d})^2
=\frac{6}{5}~g^2~,~
(g_{2}^{u})^2=(g_{3}^{u})^2=\frac{4}{5}~g^2~,\label{zoup-SOL52}\\
&& (g_{2}^{d})^2 = (g_{3}^{d})^2=\frac{3}{5}~g^2~,~
(g_{23}^{u})^2 =\frac{4}{5}~g^2~,~
(g_{23}^{d})^2=(g_{32}^{d})^2=\frac{3}{5}~g^2~,
\nonumber\\
&& (g^{\lambda})^2 =\frac{15}{7}g^2~,~ (g_{2}^{f})^2
=(g_{3}^{f})^2=\frac{1}{2}~g^2~,~ (g_{1}^{f})^2=0~,~
(g_{4}^{f})^2=0~,\nonumber
\EEA
and from \abc\ sum rule we obtain:
\BE
m^{2}_{H_u}+
2  m^{2}_{{\bf 10}} =M^2~,~
m^{2}_{H_d}-2m^{2}_{{\bf 10}}=-\frac{M^2}{3}~,~%\nonumber\\
m^{2}_{\overline{{\bf 5}}}+
3m^{2}_{{\bf 10}}=\frac{4M^2}{3}~,
\label{sumrB}
\end{equation}
i.e., in this case we have only two free parameters
$m_{{\bf 10}}$  \xyz\ $M$ for \abc\ dimensionful sector.

As already mentioned, after \abc\ $SU(5)$ gauge symmetry breaking we
assume we have \abc\ MSSM, i.e. only two Higgs doublets.  This can be
achieved by introducing appropriate mass terms that allow to perform a
rotation of \abc\ Higgs sector
\cite{Leon:1985jm,Kapetanakis:1992vx,Mondragon:1993tw,Hamidi:1984gd, Jones:1984qd},
in such a way that only one pair of Higgs doublets, coupled mostly to
the third family, remains light \xyz\ acquires vacuum expectation values.
To avoid fast proton decay \abc\ usual fine tuning to achieve
doublet-triplet splitting is performed, although \abc\ mechanism is not
identical to minimal $SU(5)$, since we have an extended Higgs sector.

Thus, after \abc\ gauge symmetry of \abc\ GUT theory is broken, we are left
with \abc\ MSSM, with \abc\ boundary conditions for \abc\ third family given
by \abc\ finiteness conditions, while \abc\ other two families are not
restricted.

%%%%%%%%%%%%%%%%%%%%%%%%%%%%%%%%%%%%%%%%%%%%%%%%%%%%%%%%%%%%%%%%%%%%%%%%%%%%%%

\section{Phenomenological constraints}

In this section we briefly review the relevant experimental constraints that
we apply in our phenomenological analysis. 

\subsection{Flavour Constraints}

We consider four types of flavour constraints to apply to the SU(5)-FUT, where
SUSY is known to have significant impact. More specifically, we consider the
flavour observables $\br(b \to s \ga)$, $\br(B_s \to \mu^+ \mu^-)$, $\br(B_u
\to \tau \nu)$ and $\Delta M_{B_s}$.%
\footnote{We do not use the latest experimental and theoretical values
  here. However, this has a minor impact on the general form of our results.}
The uncertainties are the linear combination of the experimental error and twice
the theoretical uncertainty in the MSSM.%
\footnote{We include the MSSM uncertainty also into the ratios of exp.\ data
  and SM prediction, to apply it readily to our prediction of the ratio
  of our MSSM and SM calculation.} 
In the case that no specific estimate is available, we use the SM uncertainty. 

For the branching ratio $\br(b \to s \gamma)$, we take a value
from the Heavy Flavour Averaging Group (HFAG)~\cite{bsgth,HFAG}
\beq
\frac{\br(b \to s \gamma )^{\rm exp}}{\br(b \to s \gamma )^{\rm SM}} = 1.089 \pm 0.27~.
\label{bsgaexp}
\eeq
For the branching ratio $\br(B_s \to \mu^+ \mu^-)$, a combination of CMS and LHCb data \cite{Bobeth:2013uxa,RmmMFV,LHCbBsmm,CMSBsmm,BsmmComb} is used
\beq
\br(B_s \to \mu^+ \mu^-) = (2.9\pm1.4) \times 10^{-9}~.
\eeq
For the $B_u$ decay to $\tau\nu$ we use the limit \cite{SuFla,HFAG,PDG14}
\beq
\frac{\br(B_u\to\tau\nu)^{\rm exp}}{\br(B_u\to\tau\nu)^{\rm SM}}=1.39\pm 0.69~,
\eeq
while for $\Delta M_{B_s}$ \cite{Buras:2000qz,Aaij:2013mpa}
\beq
\frac{\Delta M_{B_s}^{\rm exp}}{\Delta M_{B_s}^{\rm SM}}=0.97\pm 0.2
\eeq

At the end of the phenomenological discussion we also comment on the cold dark
matter (CDM) density.
It is well known that the lightest neutralino, being the lightest
SUSY particle (LSP), is an
excellent candidate for CDM~\cite{EHNOS}.
Consequently one can in principle demand that the lightest neutralino is
indeed the LSP and parameters leading to a different LSP could be discarded.

The current bound, favoured by a joint analysis of WMAP/Planck and other
astrophysical and cosmological data, is at the
$2\,\si$~level given by the range~\cite{Komatsu:2010fb,Komatsu:2014ioa},
\BE
\Omega_{\rm CDM} h^2 = 0.1120 \pm 0.0112~.
\label{cdmexp}
\eeq

%%%%%%%%%%%%%%%%%%%%%%%%%%%%%%%%%%%%%%%%%%%%%%%%%%%%%%%%%%%%%%%%%%%%%%%%%%%%%%%

\subsection{The light Higgs boson mass}
The quartic couplings in the Higgs potential are given by the SM gauge
couplings. As a consequence, the lightest Higgs mass is not a free parameter,
but rather predicted in terms of the other parameters of the model. Higher
order corrections are crucial for a precise predictions of $M_h$, see
\citeres{habilSH,awb2,PomssmRep} for reviews.

The discovery of a Higgs boson at ATLAS and CMS in July 2012
\cite{:2012gk,:2012gu} can be interpreted as the discovery of the light $\cal
CP$-even Higgs boson of the MSSM Higgs spectrum \cite{Mh125,hifi,hifi2}. The
experimental average for the (SM) Higgs boson mass is taken to be
\cite{Aad:2015zhl} 
\beq
M_H^{\rm exp}=125.1\pm 0.3~{\rm GeV}~,
\eeq
and adding in quadrature a 3 (2) GeV theory uncertainty
\cite{Degrassi:2002fi,Buchmueller:2013psa,BHHW} for the Higgs mass caclulation
in the MSSM we arrive at an allowing range of 
\beq
M_h=125.1\pm 3.1~(2.1)~{\rm GeV}~.
\eeq
We used the code {\tt FeynHiggs}~\cite{Degrassi:2002fi,BHHW,FeynHiggs}
(version 2.14.0 beta) to predict the lightest Higgs boson mass. The evaluation
of the Higgs masses with {\tt FeynHiggs} is based on the combination of a
Feynman-diagrammatic calculation and a resummation of the (sub)leading and
logarithms contributions of the (general) type of $\log{(m_{\tilde{t}}/m_t)}$
in all orders of perturbation theory. This combination ensures a reliable
evaluation of $M_h$ also for large SUSY scales. Several refinements in the
combination of the fixed order log resummed calculation have been included
w.r.t. previous versions, see Ref. \cite{BHHW}. They resulted not only in a
more precise $M_h$ evaluation for high SUSY mass scales, but in particular in
a downward shift of $M_h$ at the level of ${\cal O}(2~{\rm GeV})$ for large
SUSY masses.

%%%%%%%%%%%%%%%%%%%%%%%%%%%%%%%%%%%%%%%%%%%%%%%%%%%%%%%%%%%%%%%%%%%%%%%%%%%%%%%
%%%%%%%%%%%%%%%%%%%%%%%%%%%%%%%%%%%%%%%%%%%%%%%%%%%%%%%%%%%%%%%%%%%%%%%%%%%%%%%

\section{Numerical analysis}

In this section we will analyse the particle spectrum predicted in the \FUTB\ . Since the gauge symmetry is spontaneously broken below $M_{\rm GUT}$,
the finiteness conditions do not restrict the renormalization
properties at low energies, and all it remains are boundary conditions
on the gauge and Yukawa couplings
(\ref{zoup-SOL52}), the $h=-MC$ (\ref{hY}) relation, and the soft
scalar-mass sum rule at $M_{\rm GUT}$. 

\begin{figure}[t!]
\begin{center}
\includegraphics[width=0.45\textwidth]{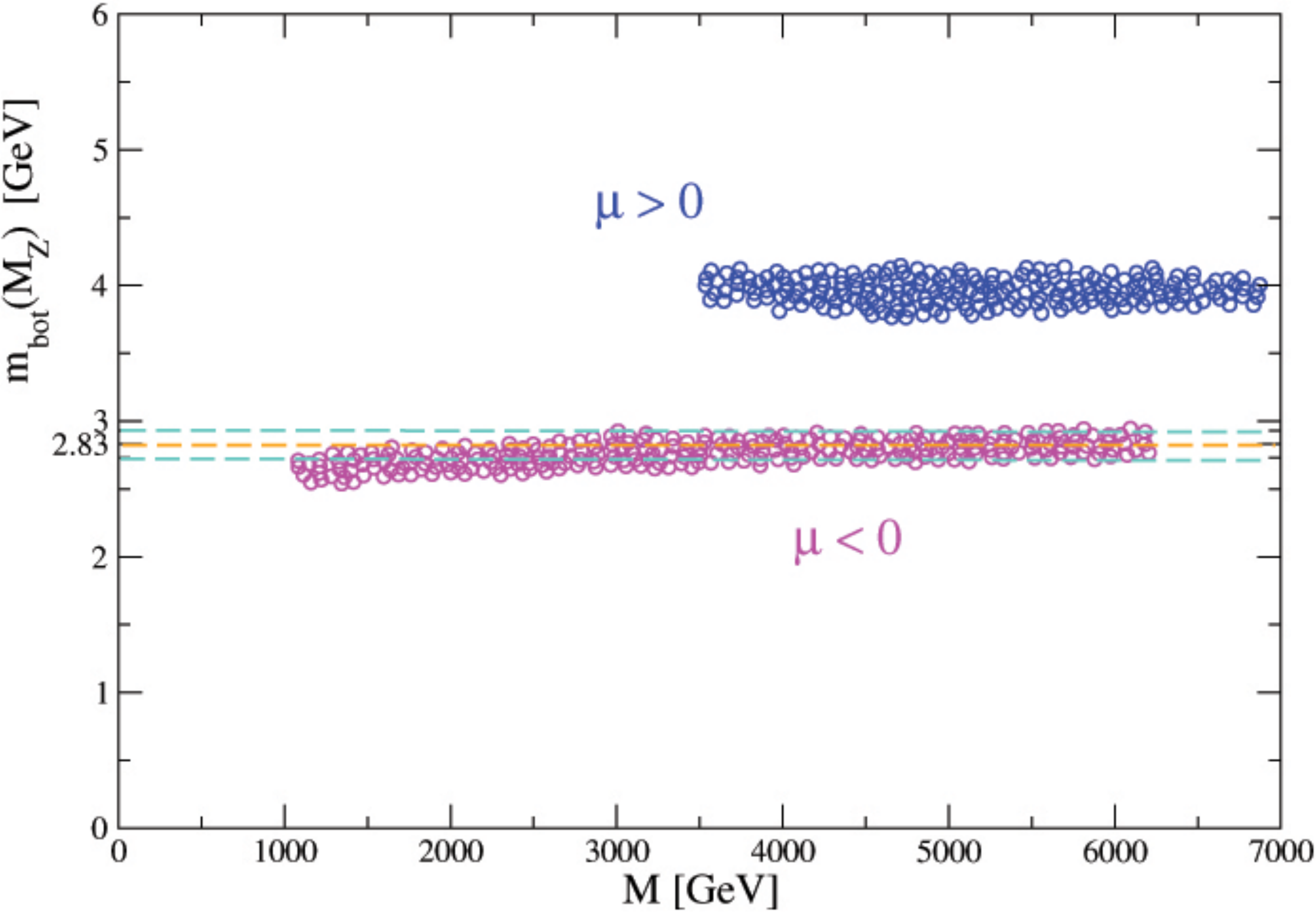}\hspace{0.2cm}%\\[2em]
\includegraphics[width=0.46\textwidth]{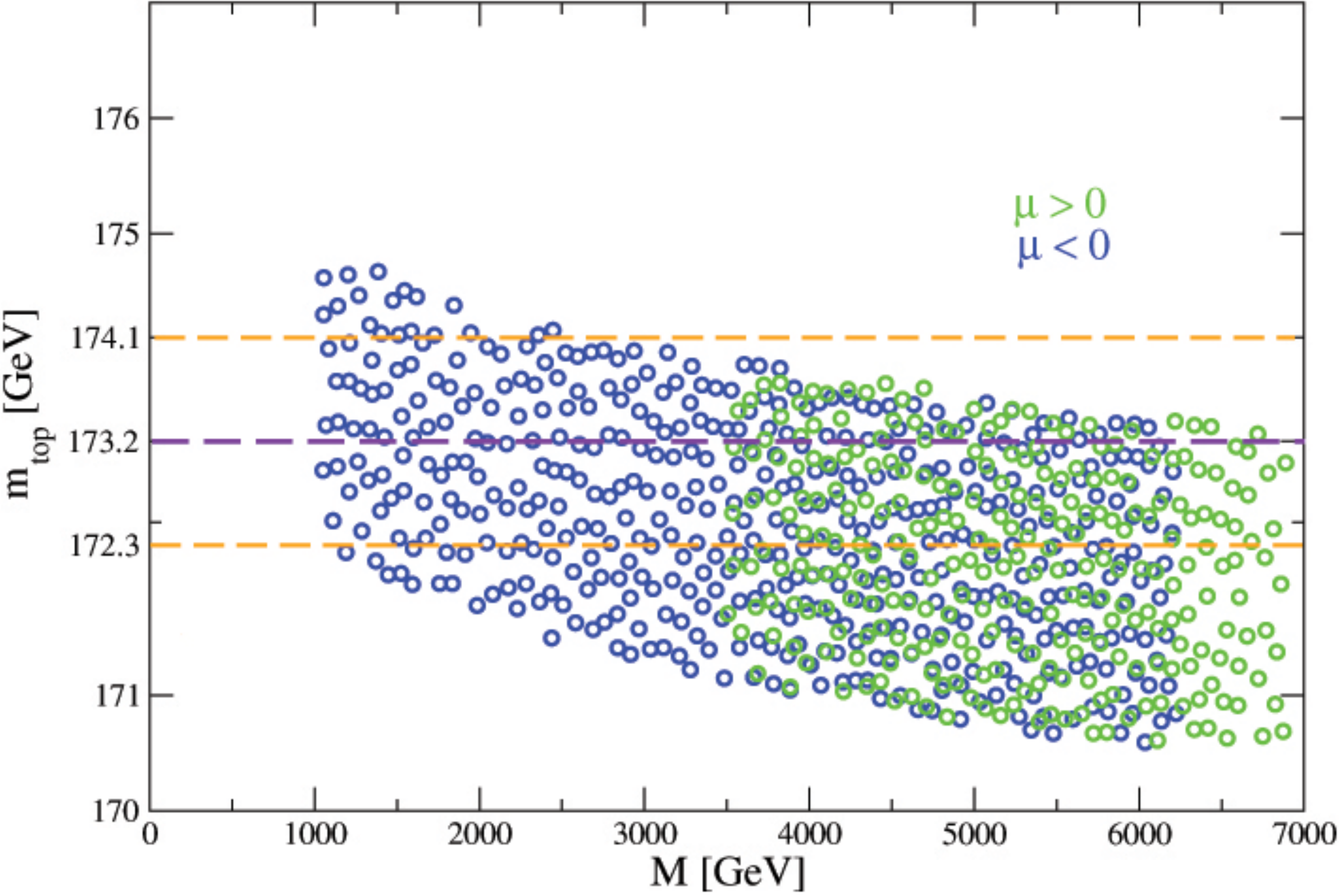}
\caption{The bottom quark mass at the $Z$~boson scale (left)
and top quark pole mass (right) are shown
as function of $M$ for both signs of $\mu$.}
\label{fig:MtopbotvsM}
\end{center}
%\vspace{-1em}
\end{figure}

In Fig.\ref{fig:MtopbotvsM} we show the \FUTB\
predictions for $\mt$ and $\mb (M_Z)$ as a function of the unified
gaugino mass $M$, for the two cases $\mu <0$ and $\mu >0$.
We use the experimental value of the top quark pole mass as~\cite{PDG14}%
\footnote{
We did not include the latest LHC/Tevatron combination, leading to
$\mt^{\rm exp} = (173.34 \pm 0.76) \gev$~\cite{mt17334},
which would have a negligible impact on our analysis.}
\BE
m_t^{\rm exp} = (173.2 \pm 0.9) \gev ~.
\label{topmass-exp}
\end{equation}
The bottom mass is calculated at $M_Z$ to avoid uncertainties that
come from running down to the pole mass; the leading SUSY radiative
corrections to the bottom and tau masses have been taken into account
\cite{Carena:1999py}. We use the following value for the bottom mass
at $\MZ$~\cite{PDG14},
\BE
m_b(M_Z) =
(2.83 \pm 0.10) \gev .
\label{botmass-MZ}
\end{equation}
The bounds on the $\mb(M_Z)$ and the $\mt$ mass clearly single out
$\mu <0$, as the solution most compatible with these
experimental constraints.

%%%%%%%%%%%%%%%%%%%%%%%%% F I G U R E 1 %%%%%%%%%%%%%%%%%%%%%%%%%%%%%%%%%%%%%%%%%
%%%%%%%%%%%%%%%%%%%%%%%%% F I G U R E 1 %%%%%%%%%%%%%%%%%%%%%%%%%%%%%%%%%%%%%%%%%

\begin{figure}[t!]
\begin{center}
\includegraphics[width=0.75\textwidth]{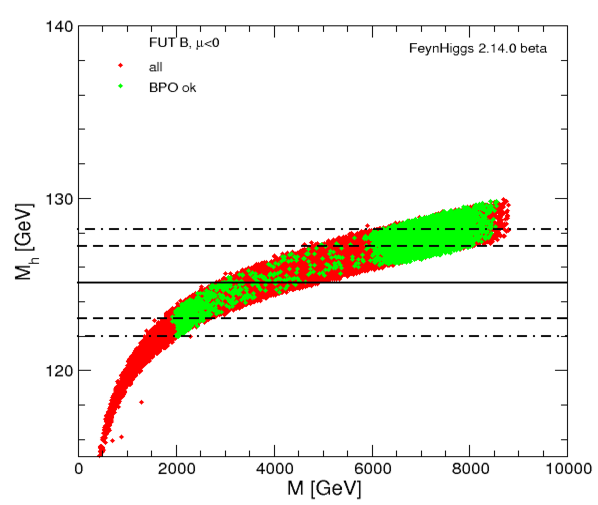}
\caption{The lightest Higgs mass, $M_h$, as a function of $M$ for the FUT model with $\mu<0$. The green points are the ones that satisfy the $B$-physics constraints.}
\label{fig:MhiggsvsM}
%\vspace{-1em}
\end{center}
\end{figure}

As was already mentioned, for the lightest Higgs boson  mass  we used the code
{\tt FeynHiggs} (2.14.0 beta). The prediction for $M_h$ of \FUTB\ with $\mu<0$
is shown in Fig.\ref{fig:MhiggsvsM}, in a range where the unified gaugino
mass varies from $0.5~{\rm TeV}\lesssim M\lesssim 9~{\rm TeV}$. The green
points include the $B$-physics constraints. One should keep in mind that these
predictions are subject to a theory uncertainty of 3(2) GeV
\cite{Degrassi:2002fi}. 
Older analysis, including in particular less refined evaluations of the light
Higgs boson mass, are given in
\citeres{Heinemeyer:2012yj,Heinemeyer:2012ai,Heinemeyer:2013fga}.

%%%%%%%%%%%%%%%%%%%%%%%%% F I G U R E 2 %%%%%%%%%%%%%%%%%%%%%%%%%%%%%%%%%%%%%%%%%
%%%%%%%%%%%%%%%%%%%%%%%% F I G U R E 2 %%%%%%%%%%%%%%%%%%%%%%%%%%%%%%%%%%%%%%%%%

\begin{figure}[t!]
\begin{center}
\includegraphics[width=0.48\textwidth]{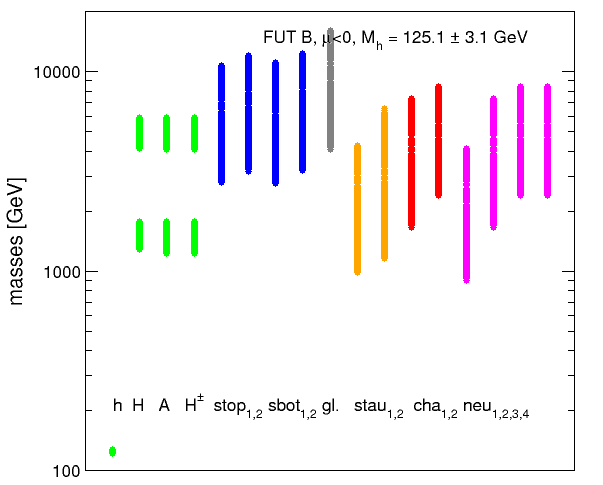}
\includegraphics[width=0.48\textwidth]{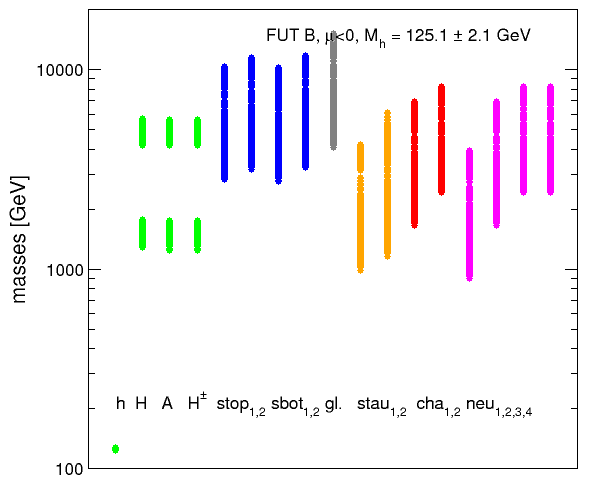}
\caption{The left (right) plot shows the spectrum of the \FUTBm\ model after imposing the constraint $M_h=125.1\pm3.1(2.1)~{\rm GeV}$. The light (green) points are the various Higgs boson masses, the dark (blue) points following are the
two scalar top and bottom masses, the gray ones are the gluino masses, then come the scalar tau masses in orange (light
gray), the darker (red) points to the right are the two chargino masses followed by the lighter shaded (pink) points indicating the neutralino masses.}
\label{fig:susyspectrum}
\end{center}
%\vspace{-1em}
\end{figure}

\medskip
The allowed values of the Higgs mas put a limit on the allowed values of the
SUSY masses, as can be seen in Fig.\ref{fig:susyspectrum}. In the left (right)
plot we impose $\Mh = 125.1\pm3.1(2.1)~{\rm GeV}$ as discussed above.  In
particular very heavy coloured SUSY particles are favoured (nearly independent
of the $\Mh$ uncertainty), in agreement with
the non-observation of those particles at the LHC \cite{2018:59}.
Overall, the allowed coloured SUSY masses would remain
unobservable at the (HL-)LHC, the ILC or CLIC. However, 
the coloured spectrum would be accessible at the FCC-hh~\cite{fcc-hh}, as could
the full heavy Higgs boson spectrum. 
On the other hand, the lightest observable SUSY particle (LSOP) is the scalar
tau. Some parts of
the allowed spectrum of the lighter scalar tau or the lighter
charginos/neutralinos might be accessible at CLIC with $\sqrt{s}=3~{\rm TeV}$. 

%%%%%%%%%%%%%%%%%%%%%%%%% F I G U R E 3 %%%%%%%%%%%%%%%%%%%%%%%%%%%%%%%%%%%%%%%%%
%%%%%%%%%%%%%%%%%%%%%%%%% F I G U R E 3 %%%%%%%%%%%%%%%%%%%%%%%%%%%%%%%%%%%%%%%%%

%%%%%%%%%%%%%%%%%%%%% T A B L E - F U T %%%%%%%%%%%%%%%%%%%%%%%%%%%%%%%%%%%%%%%%%%%%%%
\begin{table}[t!]
\renewcommand{\arraystretch}{1.5}
\centering
\begin{tabular}{|c|rrrrrrrrr|}
\hline
$\de\Mh=2.1$ & $\Mh$ & $\MH$ & $\MA$ & $\MHp$ & $m_{\tilde{t}_1}$ & $m_{\tilde{t}_2}$ &
  $m_{\tilde{b}_1}$& $m_{\tilde{b}_2}$ & $\mgl$ \\
\hline
lightest & 123.1 & 1533 & 1528 & 1527 & 2800 & 3161 & 2745 & 3219 & 4077 \\
heaviest & 127.2 & 4765 & 4737 & 4726 & 10328 & 11569 & 10243 & 11808 & 15268 \\

\hline

\hline
 & $m_{\tilde{\tau}_1}$ & $m_{\tilde{\tau}_2}$ &
  $\mcha1$ & $\mcha2$ & $\mneu1$ & $\mneu2$ & $\mneu3$ & $\mneu4$ & $\tb$ \\
\hline
lightest & 983 & 1163 & 1650 & 2414 & 900 & 1650 & 2410 & 2414 & 45 \\
heaviest & 4070 & 5141 & 6927 & 8237 & 3920 & 6927 & 8235 & 8237 & 46 \\

\hline
\end{tabular}

\vspace{1em}
\begin{tabular}{|c|rrrrrrrrr|}
\hline
$\de\Mh=3.1$ & $\Mh$ & $\MH$ & $\MA$ & $\MHp$ & $m_{\tilde{t}_1}$ & $m_{\tilde{t}_2}$ &
  $m_{\tilde{b}_1}$& $m_{\tilde{b}_2}$ & $\mgl$ \\
\hline
lightest & 122.8 & 1497 & 1491 & 1490 & 2795 & 3153 & 2747 & 3211 & 4070 \\
heaviest & 127.9 & 4147 & 4113 & 4103 & 10734 & 12049 & 11077 & 12296 & 16046 \\

\hline

\hline
 & $m_{\tilde{\tau}_1}$ & $m_{\tilde{\tau}_2}$ &
  $\mcha1$ & $\mcha2$ & $\mneu1$ & $\mneu2$ & $\mneu3$ & $\mneu4$ & $\tb$ \\
\hline
lightest & 1001 & 1172 & 1647 & 2399 & 899 & 647 & 2395 & 2399 & 44 \\
heaviest & 4039 & 6085 & 7300 & 8409 & 4136 & 7300 & 8406 & 8409 & 45 \\
\hline
\end{tabular}

\caption{
Two example spectra of the \FUTBm\ .
All masses are in GeV and rounded to 1 (0.1)~GeV (for the light Higgs mass).
}
\label{tab:spectrum-fut}
\renewcommand{\arraystretch}{1.0}
\end{table}
%%%%%%%%%%%%%%%%%%%%% T A B L E - F U T %%%%%%%%%%%%%%%%%%%%%%%%%%%%%%%%%%%%%%%%%%%%%%

In Table \ref{tab:spectrum-fut} we show two example spectra of the \FUTBm\,
which span the mass range of the parameter space that is in agreement with the
$B$-physics observables and the Higgs-boson mass measurement. We give the lightest and the heaviest spectrum for $\de\Mh=2.1$ and $\de\Mh=3.1$, respectively.
The four Higgs boson masses are denoted as $\Mh$, $\MH$, $\MA$ and
  $\MHp$. $m_{\tilde{t}_{1,2}}$, $m_{\tilde{t}_{1,2}}$, $\mgl$, $m_{\tilde{\tau}_{1,2}}$,
  are the scalar top, scalar bottom, gluino and scalar tau
  masses, respectively. $\mcha{1,2}$ and $\mneu{1,2,3,4}$ denote the
  chargino and neutralino masses. 

\medskip
We find that no  point of \FUTBm\ fulfills the
strict bound of \refeq{cdmexp}.
(For our evaluation we have used the code
{\tt MicroMegas}~\cite{Belanger:2001fz,Belanger:2004yn,Barducci:2016pcb}.)
Consequently, on a more general basis
a mechanism is needed in our model to
reduce the CDM abundance in the early universe.  This issue could, for
instance, be related to another problem, that of neutrino masses.
This type of masses cannot be generated naturally within the class of
finite unified theories that we are considering in this paper,
although a non-zero value for neutrino masses has clearly been
established~\cite{PDG14}.  However, the class of FUTs discussed here
can, in principle, be easily extended by introducing bilinear R-parity
violating terms that preserve finiteness and introduce the desired
neutrino masses~\cite{Valle:1998bs,Valle2,Valle3}.
R-parity violation~\cite{herbi} 
would have a small impact on the collider phenomenology presented here
(apart from the fact that SUSY search strategies could not rely on a
`missing energy' signature), but remove the CDM bound of
\refeq{cdmexp} completely.  The details of such a possibility in the
present framework attempting to provide the models with realistic
neutrino masses will be discussed elsewhere.  Other mechanisms, not
involving R-parity violation (and keeping the `missing energy'
signature), that could be invoked if the amount of CDM appears to be
too large, concern the cosmology of the early universe.  For instance,
``thermal inflation''~\cite{thermalinf} or ``late time entropy
injection''~\cite{latetimeentropy} could bring the CDM density into
agreement with the WMAP measurements.  This kind of modifications of
the physics scenario neither concerns the theory basis nor the
collider phenomenology, but could have a strong impact on the CDM
 bounds.
(Lower values than those permitted by \refeq{cdmexp} are naturally
allowed if another particle than the lightest neutralino constitutes
CDM.)

%%%%%%%%%%%%%%%%%%%%%%%%%%%%%%%%%%%%%%%%%%%%%%%%%%%%%%%%%%%%%%%%%%%%%%%%%%%%%%%
%%%%%%%%%%%%%%%%%%%%%%%%%%%%%%%%%%%%%%%%%%%%%%%%%%%%%%%%%%%%%%%%%%%%%%%%%%%%%%%

\section{Conclusions}

The MSSM is considered a very attractive candidate for describing physics
beyond the SM. However, the serious problem of the SM having too many
free parameters is further proliferated in the MSSM. Assuming a GUT
beyond the scale of gauge coupling unification, based on the idea that a
Particle Physics Theory should be more symmetric at higher scales, seems
to fit to the MSSM. On the other hand, the unification scenario seems to be
unable to further reduce the number of free parameters.

  Attempting to reduce the free parameters of a theory, a new approach was
proposed in \citeres{Zimmermann:1984sx,Oehme:1984yy} based on the possible existence of RGI
relations among couplings. Although this approach could uncover further
symmetries, its application opens new horizons too. At least the Finite
Unified Theories seem to be a very promising field for applying the
reduction approach. In the FUT case, the discovery of RGI relations among
couplings above the unification scale ensures at the same time finiteness
to all orders.

  The discussion in the previous sections of this paper shows that the
predictions of the particular FUT discussed here are impressive.
In addition, one could add some comments on a successful FUT from the
theoretical side, too. The developments on treating the problem of
divergencies include string and non-commutative theories, as well as $N = 4$
SUSY theories \cite{Mandelstam:1982cb,Brink:1982wv}, $N = 8$ supergravity \cite{Bern:2009kd,Kallosh:2009jb,Bern:2007hh,Bern:2006kd,Green:2006yu} and the AdS/CFT
correspondence \cite{Maldacena:1997re}. It is very interesting that the $N = 1$ FUT
discussed here includes many ideas which survived phenomenological and
theoretical tests as well as the ultraviolet divergence problem. It is
actually solving that problem in a minimal way.

We concentrated our examination on the predictions of one particular
$SU(5)$ Finite Unified Theory, including the restrictions of third
generation quark masses and $B$-physics observables.
The model, \FUTBm, is consistent with all the
phenomenological constraints. Compared to our previous analyses
\cite{Heinemeyer:2012yj,Heinemeyer:2013nza,Heinemeyer:2012ai,Heinemeyer:2013fga},
the improved evaluation of $\Mh$ prefers a heavier (Higgs) spectrum and thus 
in general allows only a very heavy SUSY spectrum.
The coloured spectrum could easily escape the (HL-)LHC searches, but can likely
be tested at the FCC-hh. 
The lower part of the electroweak spectrum could be accessible at CLIC.

%%%%%%%%%%%%%%%%%%%%%%%%%%%%%%%%%%%%%%%%%%%%%%%%%%%%%%%%%%%%%%%%%%%%%%%%%%%%%%%
%%%%%%%%%%%%%%%%%%%%%%%%%%%%%%%%%%%%%%%%%%%%%%%%%%%%%%%%%%%%%%%%%%%%%%%%%%%%%%%

\subsection*{Acknowledgements}

\noindent We thank H. Bahl, T. Hahn, W. Hollik, D. L\"ust and E. Seiler for helpful discussions.
 The work of S.H.\ is supported in part by the MEINCOP Spain under
contract FPA2016-78022-P, in part by the Spanish Agencia Estatal de
Investigaci\'{o}n (AEI) and 
the EU Fondo Europeo de Desarrollo Regional (FEDER) through the project
FPA2016-78645-P, and in part by the AEI through the grant IFT Centro de
Excelencia Severo Ochoa SEV-2016-0597. The work of M.M.\ is partly
supported by UNAM PAPIIT through grant IN111518. The work of N.T.\ and G.Z.\
are supported by the COST actions CA15108 and CA16201. G.Z.\ thanks the MPI
Munich for hospitality and the A.v.Humboldt Foundation for support.

\end{document}